\documentclass{elsart}
\usepackage{epsfig}
\begin{document}
\runauthor{Wendeker {\em et al.}}

\begin{frontmatter}
\title{Chaotic Combustion in Spark Ignition Engines}
\author{Miros\l{}aw Wendeker\thanksref{E-mail}}
\author{, Jacek Czarnigowski}
\author{, Grzegorz Litak\thanksref{E-mail1}} 
\author{and Kazimierz Szabelski}  
\address{Department of Mechanics, Technical University of Lublin, 
Nadbystrzycka 36, PL-20-618 Lublin, Poland}

\thanks[E-mail]{Fax: +48-815250808; E-mail:
wendeker@archimedes.pol.lublin.pl}
\thanks[E-mail1]{Fax: +48-815250808; E-mail:
litak@archimedes.pol.lublin.pl}

\begin{abstract}
We analyse the combustion process in an spark
ignition  engine  using
 the experimental data of an internal pressure during  the combustion process and
show that the system can be driven to chaotic behaviour. Our conclusion is based on
the observation of unperiodicity in the time series, suitable stroboscopic maps and
a complex structure of a reconstructed strange attractor.  This analysis can explain
that in some circumstances the level of noise in spark ignition engines increases considerably
due to nonlinear dynamics of a combustion process.  
\end{abstract}
\end{frontmatter}

\section{Introduction}

It is known that the cyclic combustion variability is one of the main characteristics for spark 
ignition (SI) engines. If cyclic variability were eliminated, there would be even 10\% increase
in the power output  of the engine \cite{Hu96}. Cyclic variations in the combustion process of SI
engines have been a subject of an intensive research in  last 40 years. 
Heywood \cite{Hey88} identified 
three main factors influencing cycle-to-cycle variations: 
aerodynamic in the cylinder during 
combustion, the amount of fuel, air and recycled exhaust 
gases supplied to the cylinder and a 
mixture composition near the spark plug.

Recently, in the context of engine control,  there 
appeared a lot of papers on dynamic phenomenon identifications and predictions of engine 
behaviour
\cite{Daw96,Rob97,Wen99,Wen02,Ant02}.
Many of them discuss the problem of a high noise level which make the engine difficult 
to control. 
On this basis several stochastic models have been applied to  pressure cyclic 
variation analyzing \cite{Rob97,Wen99}. 

As the combustion of gases in the cylinder may be viewed as a nonlinear dynamical process 
\cite{Hu96,Daw96}, it may
be studied by tools belonging to the theory of nonlinear systems. The variations might 
originate
in nonlinear dependence of the peak cycle temperature and pressures on the 
initial conditions at the
beginning of compression due to exhaust recirculation and a mixture  preparation process.
Such nonlinear approach has been already started by Daw {\em et al.} \cite{Daw96} 
where he suggested 
the possible chaotic nature of combustion. Our paper explanation is going 
in the same direction, 
however in our approach we adopted the novel measurement method basing on optical fibers to measure 
the pressure directly \cite{Cza02}.

\section{Chaotic oscillations of internal pressure}

We start our analysis monitoring the measured time series for various spark advanced angles $\Delta 
\alpha_z$. 
The crankshaft frequency was chosen to be
$\Theta=14.17$ Hz  ($\Theta=850$ RPM) as an idle speed of the engine. 
In a combustion process only one of each two cycles
of
crankshaft rotation
are coincided with combustion   
leading to pressure oscillation with a period $2/\Theta=0.141$s.  
Note that depending on  ignition timing 
($\Delta \alpha_z$)  the system behaves 
differently. 
Starting  with $\Delta \alpha_z=5^o$ (Fig. \ref{rys1}a)  we observe 
initial periodic  oscillation.  
Increasing $\Delta \alpha_z$ to 20$^o$ the situation does not 
change very much nevertheless, one can note  that in the background of a periodic motion some 
singular instabilities occur (i.e. at t=2.8s at Fig. \ref{rys1}b) after a long interval of a stable 
motion (Fig. \ref{rys1}b).  Interestingly, going 
further to a larger value of the spark advanced angle $\Delta \alpha_z=30^o$ changes the system 
behaviour completely. 
The corresponding time history is plotted in Fig. \ref{rys1}c. In this case the system has lost 
its initial periodic nature undergoing to a more complicated type of motion.
In the last panel (Fig. \ref{rys1}d) we show the  power spectrum for this interesting case. 
The continues spectrum indicates that system can be in a chaotic state.

To clarify our conjecture about chaotic combustion we decided to present results on  the 
stroboscopic map. In this aim we prepared Fig. \ref{rys2} where we plotted the 
internal pressure coincided with the position of the crankshaft rotation angle. For enough log time 
history we see a kind of collection of Poincare maps plotted simultaneously for large number of 
performed measurements. Namely, our crankshaft was divided into 512 angles. These gives 1024 
measurement points per two cycles of a crankshaft rotation corresponding to one cycle of combustion.    
One can see that 
initially seen periodic motion (Fig. \ref{rys1}a) has already some modulation of combustion slope.
The critical region of modulation was however located behind the maximum pressure peak leaving its 
value 
$p_{max}(\phi)$
practically unaffected. Obviously,  this modulation, rather difficult to observe in Fig. \ref{rys1}a,  
introduces 
another time scale with longer period. Note that for larger value of $\Delta \alpha_z$ ($\Delta 
\alpha_z=20^o$) that critical region has moved towards the maximum  pressure peak giving rise 
to 
observable changes of the maximum pressure value (Fig. \ref{rys2}a). These changes become the most 
dramatic for $\Delta 
\alpha_z=30^o$ where the the modulations of pressure exceeds its average value (Fig. \ref{rys2}c).   
To illustrate the final effect of pressure instabilities we show, in Fig. \ref{rys2}d  maxima of 
pressure $p_{max}$ in 
sequential
cycles $n$. In this figure squares correspond to  $\Delta \alpha_z=20^o$ and circles
to  $30^o$, respectively. Note that the maximum value of pressure $p_{max}$ for $\Delta \alpha_z=5^o$ 
would be a
constant 
line in the 
scale of Fig. \ref{rys2}d.

To this end we plot in Fig. \ref{rys3} the reconstruction of the chaotic attractor.
Changing the 
parameter $\Delta \alpha_z$
which can be regarded as a bifurcation parameter we see that the attractor which was 
in principle two dimensional enlarge its dimensionality with increasing $\Delta \alpha_z$. 
The expected minimal embedding dimension was chosen arbitrary to be equal to 3 as on the basis  
that it is the smallest dimensionality enabling chaotic solutions. The characteristic time 
delay value $\tau=0.138$s was chosen simply as a number  smaller than the characteristic period
in the system $2/\Theta=0.141$s (Fig \ref{rys1}a-c). Note $2/\Theta$ as a period of a parameteric 
excitation is uneffected by 
changes of the bifurcation 
parameter 
$\Delta 
\alpha_z$.      
Starting with basically 2-d structure of the attractor in tree dimensions (Fig. \ref{rys3}a) one 
can follow  
small changes 
of it for $\Delta
\alpha_z=20^o$ (in Fig. \ref{rys3}b) and a large qualitative change in Fig. \ref{rys3}c for $\Delta
\alpha_z=30^o$.

\section{Conclusions}

In this paper we examined the oscillations of combustion internal pressure. It appeared that, 
in some conditions, intermittency is capable to  drive the system  into chaotic region. 
 We have shown that the change of  a spark advance angle  $\Delta \alpha_z$
is making  a significant effect on the combustion dynamics. It is clear that it influences  
directly
the 
flame 
initiation phase which occurs on a larger time-rate for an increasing spark advance.   
In this way the cyclic variations become an inherent phenomenon in time and in space.    
Chaotic nature of combustion in SI 
engines appears in strong sensitivity on such conditions. 
It is worth to note that Figs. \ref{rys1}c and \ref{rys1}d, for $\Delta \alpha_z=30^o$ can be easily 
associated with a chaotic 
process, but  Figs. \ref{rys1}b and \ref{rys2}b show 
that even for the advance angle  $\Delta \alpha_z=20^o$ main instabilities  
appear. This fact leads us to conclusion  that the mechanism of transition to chaos should be based 
on the intermittency phenomenon \cite{Pom80,Cha96,Lit02}.

Our preliminary 
results indicate that the noise SI engines ought to be connected with nonlinear dynamics.
Moreover analysing the combustion process for different  system parameters
we observed similar behaviour.      
I should be noted, however that  in parametric systems with an additional self-excitation the
problem of distinguishing 
between chaotic and multi frequency quasi-periodic motions is not an easy task \cite{Lit99}.
To tell more about dynamics of that particular combustion method one have to use more sophisticated 
nonlinear methods \cite{Aba96}. We plan to conduct such 
analysis in a future report \cite{Wen03}.

\begin{figure}
\section*{Figure Captions}
\caption{\label{rys1} Time series of pressure $p(t)$ in a 
combustion process
the crankshaft frequency $\Theta=850$ RPM (a-c) for various spark advance angles $\Delta \alpha_z$.
The power spectrum
for $\Delta \alpha_z$=30$^o$ (d).}
\caption{\label{rys2}  Stroboscopic projection of  internal pressure 
$p(t)$ on a relative
position of a 
crankshaft in
a combustion process $\phi$ for various $\Delta \alpha_z$ (a-c). Maxima of pressure $p_{max}$ in
sequential
pressure cycles $n$ where squares correspond to  $\Delta \alpha_z=20^o$ and circles
to  $30^o$, respectively
(d) (squares are connected by a dotted line for better clarity).}
\caption{\label{rys3} Reconstruction of the attractor for a 
combustion process in a SI engine.
Characteristic time delay was assumed to be $\tau=0.138$ s.}
\end{figure}


\begin{thebibliography}{99}
{\small
\bibitem{Hu96}
  Hu Z. Nonliner instabilities of combustion 
processes and cycle-to-cycle variations
in spark-ignition engines. SAE Technical Paper 1996:961197.
\bibitem{Hey88}  Heywood JB. Internal combustion engine 
fundamentals. New York: McGraw-Hill; 
1988. 
\bibitem{Daw96}   Daw CS, Finney CEA, Green JB Jr, Kennel MB, Thomas 
JF and Connolly FT.
A simple model for cyclic variations in a spark-ignition engine. SAE Technical Paper 1996;962086.
\bibitem{Rob97}   Roberts JB, Peyton-Jones JC, Landsborough KJ. 
Cylinder pressure variations as a
stochastic process. SAE Technical Paper 1997;970059.
\bibitem{Wen99}   Wendeker M,  Niewczas A, Hawryluk B. A stochastic 
model of the fuel injection of the
si engine.  SAE Technical Paper 1999:00P-172.
\bibitem{Wen02}    Wendeker M, Czarnigowski J. Hybrid air/fuel 
control using the adaptive estimation 
and neural network. SAE Technical Paper 2000;2000-01-1248. 
\bibitem{Ant02}   Antoni I, Daniere J, Guillet F. Effective 
vibration analysis of ic engines using 
cyclostationaryrity. Part II -- new results on the reconstruction of the cylinder pressures.
J Sound Vibr 2002;257:839-856.     
\bibitem{Cza02}   Czarnigowski J. PhD thesis, Technical University 
of Lublin 2002.
\bibitem{Pom80}   Pomeau Y, Manneville P. Intermittent transition to turbulence 
in dissipative systems.
Commun Math Phys 1980;74:189.
\bibitem{Cha96}   Chatterjee S, Malik AK. Three kinds of intermittency in a 
nonlinear system. Phys Rev E 
1996;53:4362-7. 
\bibitem{Lit02}   Litak G. Chaotic vibrations in a regenerative cutting process. 
Chaos Solitons \& 
Fractals 2002;13;1531-35. 
\bibitem{Lit99}   Litak G, Spuz-Szpoz G, Szabelski K, Warminski J. Vibration 
analysis of self--excited 
system with parametric forcing and nonlinear stiffness. Int. J. Bifurcation and Chaos 
1999;49;493-504.
\bibitem{Aba96}   Abarbanel HDI. Analysis of observed chaotic data. New 
York: 
Springer-Verlag; 1996.
\bibitem{Wen03}   Wendeker M, Czarnigowski J, Litak G, Szabelski K, to be 
published.
}
\end{thebibliography}
\end{document}